\documentclass[aps,pra,twocolumn,letterpaper,superscriptaddress,footinbib,showpacs]{revtex4-1}
\usepackage{times,graphicx,mathtools,amssymb,array,sidecap}
\usepackage[hyperfigures,colorlinks=true,allcolors=blue]{hyperref}
\usepackage[utf8]{inputenc}

\usepackage{amsmath}
\usepackage{xfrac}
\usepackage{color}

\begin{document}


\title{Self-replication with magnetic dipolar colloids}

\author{Joshua M Dempster}
\affiliation{Department of Physics and Astronomy, Northwestern University, Evanston, IL,60208}
\author{ Rui Zhang}
\affiliation{Department of Materials Science and Engineering, Northwestern University, Evanston, IL, 60208}
\author{Monica Olvera de la Cruz}
\affiliation{Department of Physics and Astronomy, Northwestern University, Evanston, IL,60208}
\affiliation{Department of Materials Science and Engineering, Northwestern University, Evanston, IL, 60208}
\affiliation{Department of Chemistry, Northwestern University, Evanston, IL, 60208}

\begin{abstract}Colloidal self-replication represents an exciting research frontier in soft matter physics. Currently, all reported self-replication schemes involve coating colloidal particles with stimuli-responsive molecules to allow switchable interactions. In this paper, we introduce a scheme using ferromagnetic dipolar colloids and pre-programmed external magnetic fields to create an autonomous self-replication system. Interparticle dipole-dipole forces and periodically varying weak-strong magnetic fields cooperate to drive colloid monomers from the solute onto templates, bind them into replicas, and dissolve template complexes. We present three general design principles for autonomous linear replicators, derived from a focused study of a minimalist sphere-dimer magnetic system in which single binding sites allow formation of dimeric templates. We show via statistical models and computer simulations that our system exhibits nonlinear growth of templates and produces nearly exponential growth (low error rate) upon adding an optimized competing electrostatic potential. We devise experimental strategies for constructing the required magnetic colloids based on documented laboratory techniques. We also present qualitative ideas about building more complex self-replicating structures utilizing magnetic colloids.      

\end{abstract}

\maketitle

\section{Introduction}
While self-replication has been extensively explored in artificial molecular systems including nucleic acids\cite{1986paper,joyce,rnaenzyme}, peptides\cite{jacs2002}, and small organic compounds\cite{rebek,vidonne}), efforts to introduce this exciting behavior in colloidal replicators have begun only very recently. A single experimental scheme was reported in 2011 which successfully reproduces template sequences of DNA-functionalized colloids using laboratory intervention at each step of the replication process\cite{chaikinsm, chaikin1}. Since then, several theoretical papers have achieved autonomous, exponential self-replication of colloids in computer simulations. However, theoretical work to date proposes coating colloids with sophisticated patterns of stimuli-responsive molecules which are not yet physically accessible. These take the form of complementary patches that can be switched on and off at will\cite{us}, bonds that spontaneously dissolve when a certain number of other bonds have formed\cite{brenner}, and bonds that are artificially restricted to form only in the presence of a template\cite{angew}.  

Here we introduce a system capable of autonomous replication and exponential growth using experimentally accessible potentials. We demonstrate that a magnetically dipolar, self-replicating colloidal system can be constructed on simple building blocks, with all features within reach of current experimental techniques. We present a simple stochastic model for dimer replicators and explicitly consider the case of a finite probability for bonds to form erroneously. This examination motivates several general design principles for autonomous linear replicators. Their application to the magnetic system produces nearly pure exponential growth. As a first proof of concept, we focus on a minimal self-replicating system comprised of magnetic monomers (constituents) and dimers (templates). 

\section{The magnetic system}
\begin{figure*}
	\includegraphics[width=7.0 in]{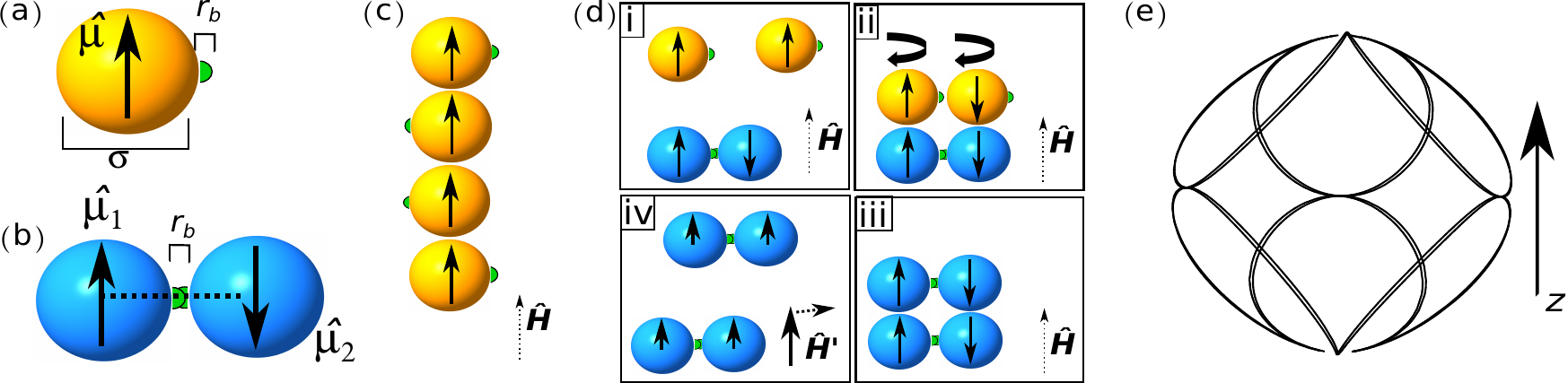}
	\caption{(Color online) The simple dipole replication scheme. (a) Detailed sketch of the dipole model used for self-replication. Binding sites (dots on the side) interact via an attractive infinite potential of range $2r_b$ with the minimum at $r_b$. Black arrows indicate magnetic dipole moment direction $\hat{\mathbf{\mu}}$. (b), A dimer template which will be replicated. If the external field strength $H$ is weak compared to $k$, the two magnetic moments in the dimer anti-align. (c) If no colloids are bound, the preferred configuration for low dipole concentrations in a weak field is a line. Binding sites are never close enough to form a new dimer. (d) The four-step replication cycle. In steps i-iii (the replication phase), a template replicates. In step iv (the mixing phase), a strong external field $\mathbf{H}'$ is reoriented continuously to dissolve clusters. (e) A sample of the direction of the mixing field $\mathbf{\hat{H}}'$ indicated by the path it traces on the unit sphere.}
	\label{fig:simple}
\end{figure*}
Both molecular\cite{1986paper,joyce,rnaenzyme,jacs2002,rebek,vidonne} and colloidal\cite{chaikin1,angew,brenner,us} self-replication follow the same basic principles. The system starts with a template of particles connected by permanent bonds. This template attracts other constituents (monomers) out of the solution and arranges them into a copy of itself. The attracted constituents are permanently bound together in a new template. Finally, new templates are separated from the old by some mechanism. At colloidal scales this requires a periodic drive, such as heat\cite{chaikin1}, light\cite{us, angew},  mechanical stress\cite{otto}, or a more elaborate internal mechanism that triggers with the formation of the new template\cite{brenner}. When templates separate and other bulk clusters dissolve, the replication cycle begins anew.

The magnetic system instantiates the general scheme just described using dipole potentials. Figure 1 describes the magnetic replication system. The system consists of a collection of hard-core spherical colloids with diameter $\sigma$ and permanent magnetic moment $\mathbf{\mu}$ fixed in each sphere's frame. All spheres are subject to the dipole-dipole interaction
\begin{equation}
	U_{12} = k \left( \frac{r_{12}}{\sigma} \right) ^{-3} \left( \hat{\mathbf{\mu}}_1 \cdot \hat{\mathbf{\mu}}_2 - 3 (\hat{\mathbf{\mu}}_1 \cdot \hat{\mathbf{r}}_{12} ) (\hat{\mathbf{\mu}}_2 \cdot \hat{\mathbf{r}}_{12} ) \right)
	\label{eq:dipole}
\end{equation}
and external potential $U_\mathrm{ext} = -\mathbf{H_{ext}} \cdot \hat{\mathbf{\mu}}$. The magnitudes of these two interactions are measured by energy parameters $k$ and $H_{ext}$. To each monomer we add one additional potential in the form of a small binding site (Fig. 1a) on the surface of the sphere, halfway between the two magnetic poles. When two binding sites move within a distance $r_b$ ($\ll \sigma$) of each other, they form a permanent, rigid bond. The resulting dimer is the template for replication (Fig. 1b). In simulations, we assume each bound dipole can still independently and freely rotate about the axis of the bond; however, to instantiate the system it is only necessary for the bond to allow 180$^\circ$ twisting from the neutral position, so that the two dipoles can switch from parallel to antiparallel. Such bonds could be created using patches of single-stranded auto-complementary DNA\cite{DNAjanus}, using one of several well-established techniques for making gold patches of controllable size\cite{janus_overview}. To ensure the patch is at the midline of the dipole, colloids should be permitted to aggregate at an interface with a field in the plane of the interface. Colloids in these interfacial aggregates will automatically present surface patches halfway between their magnetic poles. The details of one proposed interfacial fabrication method are presented in Appendix A. 

The replication sequence proceeds as follows. We begin with a random distribution of monomers and dimers in solution and allow them to evolve in a diffusive regime. An external field $\mathbf{H}$ mitigates magnetic sedimentation\cite{trusov} and accidental dimer formation by forcing monomers to aggregate only in mutually repulsive parallel strings \cite{halsey}, as illustrated in Fig. 1c. Dimers compete with these long strings for monomers. Provided $H \ll k$, dimers that succeed in attracting two monomers can trap them in a binding configuration and create a new dimer as illustrated in Fig. 1d (i-iii). We call the total time allowed for this replication phase $\tau$. After this period has elapsed, the mixing phase begins. Here the template is separated from its replica by driving the system with a strong ``mixing'' magnetic field $H' \gg k$. $\mathbf{\hat{H}}'$ changes direction much faster than the translational diffusion timescale $\tau_0$ of the monomers, sampling all orientations with a roughly equal probability. Each dipole is forced to align with the field, and dipole directions decouple from colloid displacement vectors as the field rotates. Ideally, the mean interaction between dipoles goes to 0 while the mixing field is applied:
\begin{equation}
	\langle U_{12} \rangle = k \left( \frac{r_{12}}{\sigma} \right) ^{-3} \int_{-1}^1 \mathrm{d}\cos{\theta}\,\, (1 - 3 \cos^2{\theta})=0
	\label{eq:zerodipole}
\end{equation}

In practice, a finite residual value for $\langle U_{12} \rangle < 1k_BT$ is acceptable. Once a predetermined time $t_m$ has elapsed, $H'$ is turned off and the cycle begins again with the replication phase. Continuous functions for the mixing field direction are used as shown in Fig. 1e (see more technical details in Appendix B). We make no claims that these functions are the simplest or optimal choices. Comparing alternatives is a possible avenue for future study.

\section{Growth and error in replication}
To analyze the magnetic system and assess its potential as a general self-replication method requires a basic theoretical framework, which we now present. The replication process produces new templates in proportion to existing templates and remaining monomers, at a rate we call the replication rate $R_r$: $\partial_t N_d = R_r N_d (1-2\sfrac{N_d}{N})$, where $N_d$ is the number of dimers and $N$ is the total number of monomers. However, new template bonds can generally form in the absence of templates via spontaneous collisions\cite{vidonne}. For dimer templates, this second channel is crudely proportional to the square of free monomers. We call the constant that governs this second channel the error rate $R_e$. Combining both effects leads to the total dimer growth equation:

\begin{equation}
	 \frac{\mathrm{d}N_d}{\mathrm{d}t} = \left( R_r N_d + R_e(\frac{N}{2}-N_d) \right) \left (1-2\frac{N_d}{N} \right) 
	 \label{eq:Nd_diff_full}
\end{equation}
The exact ensemble- and cycle-averaged solution for $R_r > R_e$ is 
\begin{equation}
	 N_d(t)  =\frac{N}{4(1-X)}\tanh \left ( \frac{R_r}{2}t+b \right) +\frac{N(1-2X)}{4(1-X)}
	\label{eq:Nd_full}
\end{equation}
where $X$ is the ratio $\sfrac{R_e}{R_r}$ and $b$ is chosen to match the initial number of dimers, $N_d(0)$. In the limit  $R_r \gg R_e$ and $N \gg N_d$ (equivalent to $\sfrac{1}{2}R_r t +b \ll -1$), we may simplify to
\begin{equation}
	 N_d(t) = \left ( N_d(0)+\frac{N}{2} X \right ) e^{R_rt}-\frac{N}{2} X
	\label{eq:Nd}
\end{equation} 
The two primary goals of any autonomous replication scheme are to maximize $R_r$ while minimizing $R_e$, i.e. to minimize $X$\cite{vidonne}. This is true even if the template is not a simple dimer: in that case the error channel does not produce correct templates, but instead creates malformed or mutated structures. In these more complex systems suppressing $X$ is necessary to suppress the mutation rate.

The numerator $R_e$ depends strongly on the details of the system, but a general, microscopic, stochastic model is possible for the denominator $R_r$ in the case of rigid linear templates. In the standard linear replication scheme described in the introduction, each element of the template attracts elements from the solution into a copy of the template. For this paper we consider only the creation of dimers. Let the unconditional probability of replicating a bond, and hence a dimer, in time $t$ since the start of the cycle be $P_T(t)$. With no erroneous binding and an infinite number of monomers, the number of replicated dimers for cycle $n$ will be
\begin{align}
	N_d^{(n)}=N_d^{(0)}(1+P_T(\tau))^n
	\label{eq:cyclerate}
\end{align}
As a function of time, we find
\begin{equation}
	\langle N_d(t)\rangle = N_d(0)(1+P_T(\tau))^{\frac{t}{t_m+\tau}}
	\label{eq:Nddiscrete}
\end{equation}
Equating this with $N_d(0) e^{R_rt}$ yields
\begin{equation}
	R_r(\tau)=\frac{\ln{(1+P_T(\tau))}}{t_m+\tau}
	\label{eq:Rr}
\end{equation}
in which the numerator reduces to $P_T$ for small values of $P_T$. $P_T(\tau)$ starts at zero and approaches some asymptote less than or equal to one, and therefore $R_r$ is optimized by some finite value of $\tau$. 

The unconditional probability $P_T$ can expressed as a functional of the probabilities of completing two kinds of processes: adsorbing all required monomers to the correct sites on the template, and binding the monomers into a replica. For the case of a dimer, these processes reduce to just three events:
\begin{enumerate}
\item $P_1(t_1)$, the unconditional probability of the first monomer arriving on the dimer by time $t_1$ measured from the cycle start
\item $P_2(t:t_1)$, the conditional probability of a second monomer arriving in a binding configuration by time $t>t_1$ given a first monomer arriving at $t_1$
\item $P_b(\tau:t)$, the conditional probability of a bond forming by the end of the cycle given two monomers in a binding configuration at time $t$ 
\end{enumerate}
Labeling the unconditional probability of having two monomers in a binding configuration by $t$ as $P_a(t)$, we find 

\begin{equation}
	P_a(t)=\int_0^t \mathrm{d}t_1\, \frac{\mathrm{d}P_1(t_1)}{\mathrm{d}t} P_2(t:t_1)
\end{equation}
Similarly, the unconditional probability of binding $P_T$ can be expressed as
\begin{equation}
	P_T(\tau)=\int_0^\tau \mathrm{d}t\, \frac{\mathrm{d}P_a(t)}{\mathrm{d}t} P_b(\tau:t)
	\label{eq:PT}
\end{equation}
$P_b$ must be found directly for the system under consideration either analytically or numerically. We find $P_1$ and $P_2$ by reasoning that, averaged over a large ensemble of free dimers, each dimer attracts monomers at a rate proportional to the density of free monomers, $\eta_f$. In general, monomers in the bulk form aggregates during the replication cycle so that $\eta_f$ depends on time. Therefore we find

\begin{equation}
	P_1(t_1)=1-\exp \left( -\rho \int_0^{t_1} \mathrm{d}t' \eta_f(t') \right)
	\label{eq:P1}
\end{equation}
where $\rho$ is a constant. In general the presence of the first monomer alters the rate at which monomers arrive on the neighboring site on the dimer. Let this new rate be $\xi \rho \eta_f$. If the first monomer arrives at $t_1$, the conditional solution for $P_2(t:t_1)$ is

\begin{equation}
	P_2(t:t_1)=1-\exp \left( -\xi\rho \int_{t_1}^{t} \mathrm{d}t' \eta_f(t') \right)
	\label{eq:P2}
\end{equation}

If we define a new variable $I(t) = \int_0^t \mathrm{d}s\, \eta_f(s)$, we can now write the exact probability distribution $P_a$ as

\begin{equation}
	P_a(t)=P_1(t)+\frac{e^{-\xi \rho I(t)} - e^{-\rho I(t)}}{\xi - 1}
	\label{eq:Pa_solved}
\end{equation}
and the differential form

\begin{equation}
	\frac{\mathrm{d}}{\mathrm{d}t}P_a(t) = \rho \eta_f(t)\frac{\xi}{1-\xi} \left( e^{-\rho I(t)} -  e^{-\xi \rho I(t)} \right)
	 \label{eq:dPa}
\end{equation}
We have left two free parameters, $\rho$ and $\xi$, with which to fit the model for $R_r$ to a real system. Our fitting for these parameters in the magnetic system is given in Appendix C, but the form of the dependence of $P_T$ on $I$ and $P_B$ is more significant in the following discussion.

\section{Results and Discussion}

\begin{figure*}
\centering
	\includegraphics[width=1.8\columnwidth]{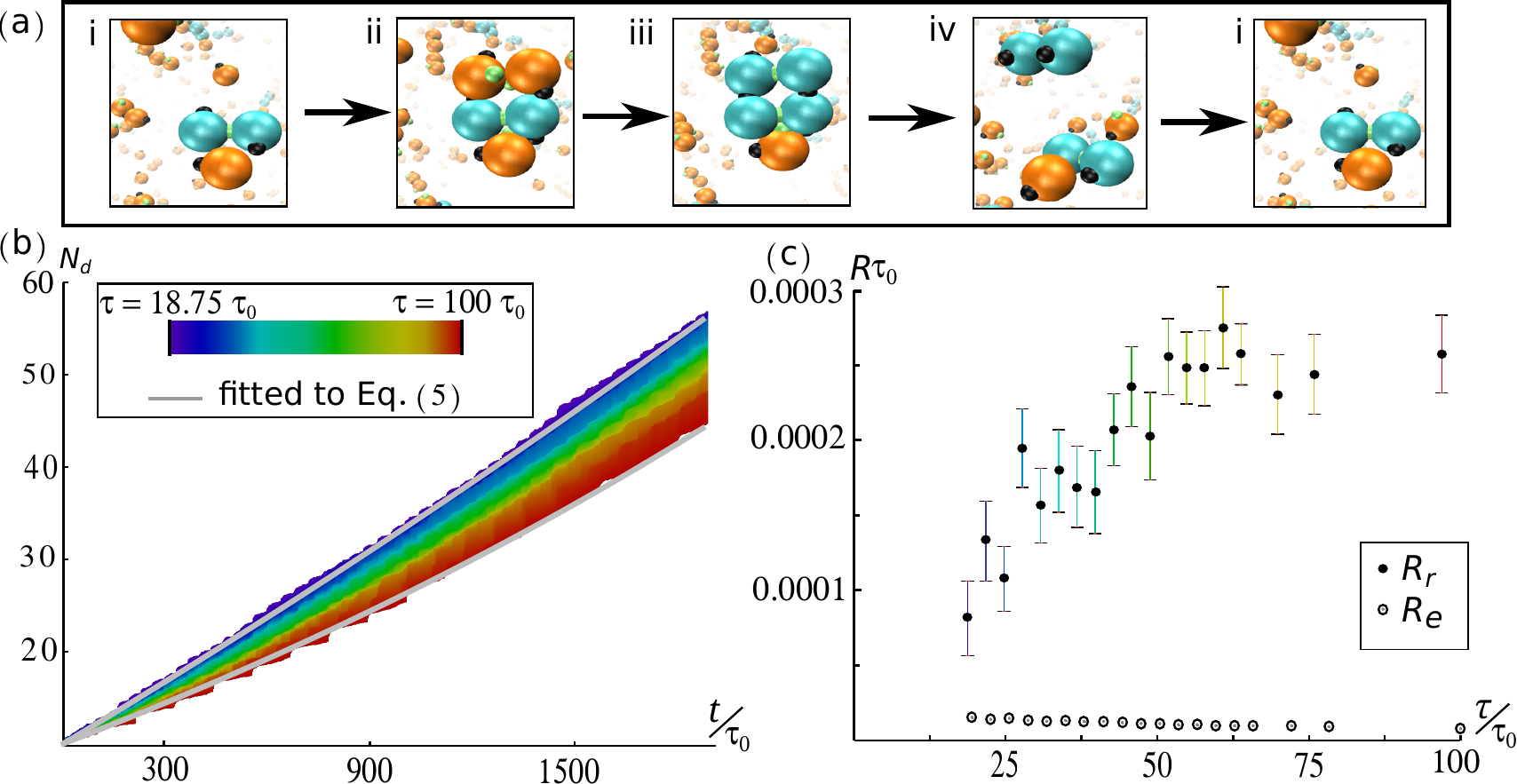}
	\caption{(Color) Dimer replication in the magnetic system.(a) Characteristic snapshots of the four steps of the replication cycle as they occur in simulations\cite{sm}. Black dots show the north pole of each dipole, and green dots show the binding sites. (b) The number of dimers $N_d$ over time for simulations with different values $\tau$. Dashed gray lines indicate best-fit models (Eq. (\ref{eq:Nd})) of dimer counts for  $\tau=25\tau_0$ and $\tau=100\tau_0$. For small values of $\tau$ the erroneous channel $R_e$ leads to entirely linear growth, while larger values begin to show non-linearity. (c) The replication rate $R_r$ and the error rate $R_e$ in units of $\tau_0^{-1}$ for various values of $\tau$. Errors were calculated using the standard deviation of best fit values, which were found using a run-replacement bootstrap method (details in Appendix C).}
	\label{fig:simple_results}
\end{figure*}

We implement a recently developed kinetic Monte Carlo simulation scheme \cite{us, jha} (a brief review is given in Appendix B) to study dimer counts as a function of time for various values of $\tau$ in systems with 2500 total colloids and 10 initial dimers. Figure 2 shows the results with $k=20$, $H=5$, $H'=1000$, $r_b=0.05\sigma$, and the volume fraction occupied by colloids $\phi=0.00226$. We confirm that growth for larger values of $\tau$ is exponential up to ternary order and that  $R_r \gg R_e$ using the statistical analysis detailed in Appendix C. However, $R_e$ still dominates dimer production unless $2\, (\sfrac{N_d}{N}) > X$. The dominant $R_e$ term causes the mostly linear growth seen. We identify $X=0.01$ as the threshold value for a replicating system to be considered nearly pure.

\begin{figure*}
\centering
	\includegraphics[width=1.8\columnwidth]{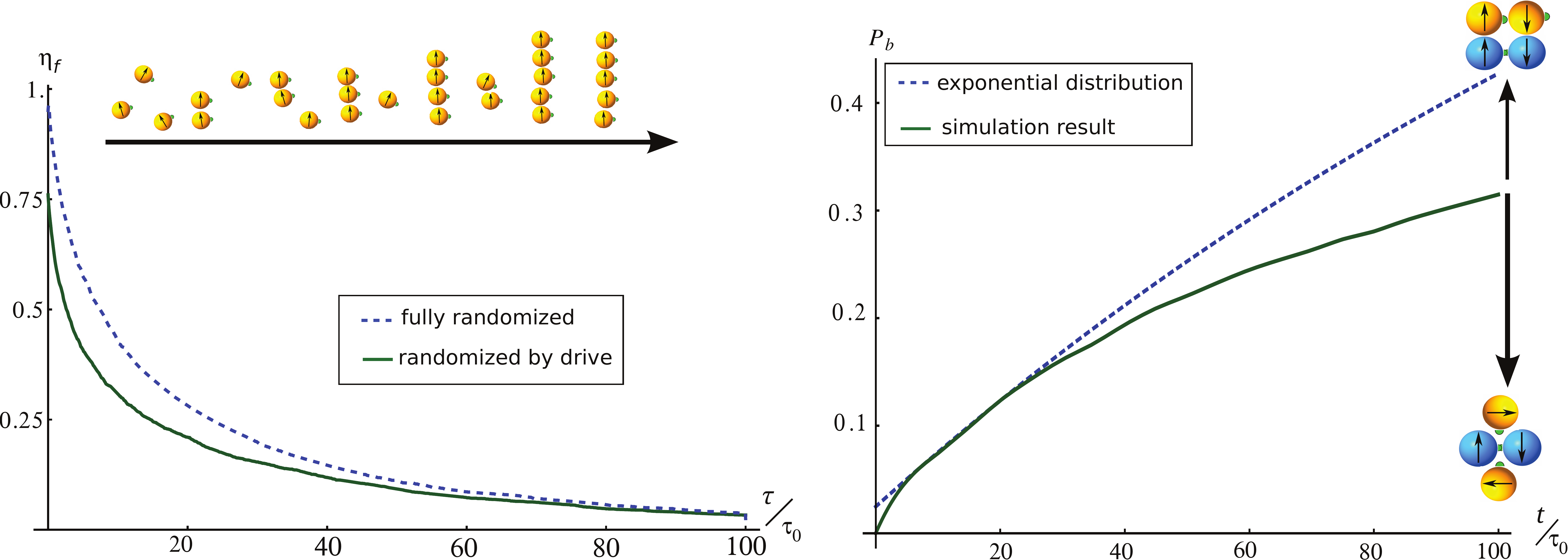}
	\caption{(Color online) The two distributions governing replication efficiency for the magnetic system.  (a) The number density of free monomers at time $t$ after the mixing cycle ends. The dotted line shows the free monomer density if the system begins in a truly random initial state, while the solid line shows the result if the system is randomized by the mixing field. Both results are averaged from 10 simulations of 500 monomers. (b) The binding probability for an isolated dimer that starts with two attracted monomers in the binding configuration at $t=0$, as in Fig. 1dii, determined by 10000 simulations of an isolated dimer/two monomer set. If the monomers remained in the binding configuration, a simple exponential distribution would be observed (shown by the dotted line). In simulations (solid line), we observe a sub-exponential saturation as monomers transition from binding to trapping configurations.}
\end{figure*}

For most replicating systems, arbitrarily small values of $X$ cannot be achieved by increasing $\tau$ alone. From eq. (\ref{eq:Rr}) is it clear that for very long values of $\tau$, the probability that any dimer replicates will saturate at a finite value $\le 1$ and $R_r$ will start falling as $\tau^{-1}$, while in the best case $R_e$ will also fall with $\tau^-1$. To reduce the lower bound of $X$ it is necessary to increase the long-$\tau$ probability $P_T(\infty)$ that dimers replicate while reducing the probability of random collisions. 

Our previous general analysis of replication provides some useful insights into achieving higher values for $P_T(\infty)$. From equations (\ref{eq:dPa}) and (\ref{eq:PT}), we find that two values limit the maximum of $P_T$. The first is $c \equiv I(\infty)$. $c$ is highest for a system that begins in a completely random state, which makes it an appropriate measure of a mixing drive's effectiveness for replication. The second limiting value is $P_B(\infty)$. Ideally, $c$ diverges and $P_B(\infty)=1$ so that it is possible to replicate every dimer with a sufficiently large value for $\tau$. In practice this is unlikely to be achieved, and one designs the drive to maximize $c$ and the energy landscape to maximize $P_B(\infty)$. Figure 3 plots $\eta_f(t)$ and $P_B(t)$ for the magnetic system with the parameters given for Fig. 2. We find that the magnetic mixing field is nearly as good at randomizing the system (maximizing $c$) as computerized randomization, with only small improvements possible. However, $P_B(t)$ saturates at values well below unity. This suggests that monomers are escaping the binding configuration and entering undesirable trapping configurations. One commonly observed trap is illustrated in Fig. 3b.

\begin{figure*}
\centering
	\includegraphics[width=1.8\columnwidth]{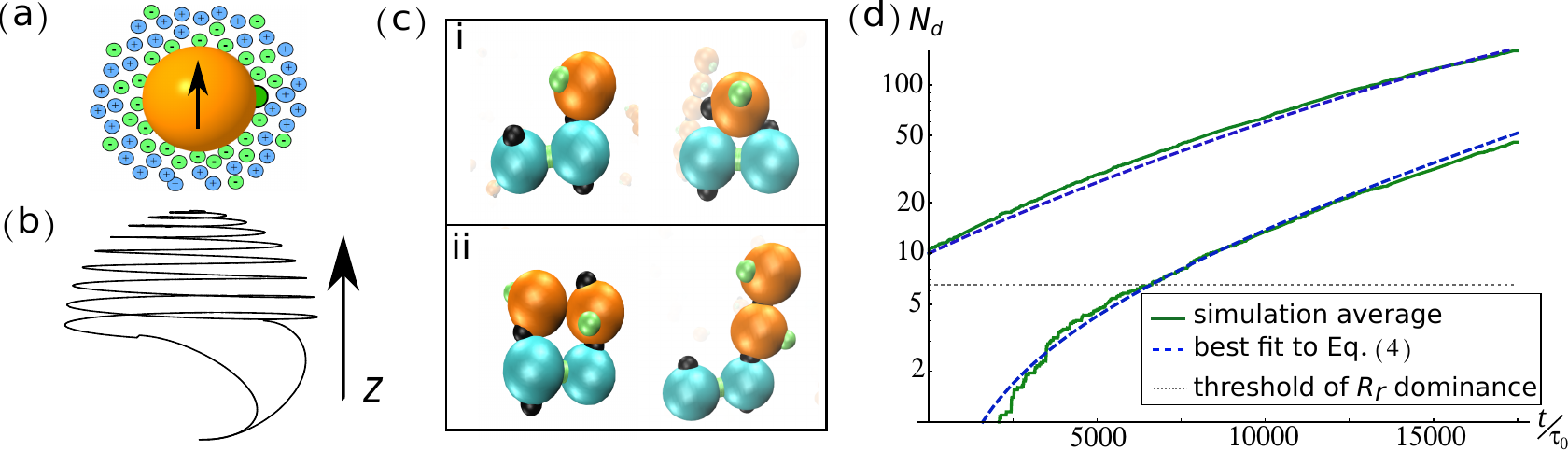}
	\caption{(Color online) Replication with electrostatic repulsion. (a) An electrically charged dipole surrounded by a double-layer of counter-ions. We choose the layer thickness to be less than $\sigma$ but greater than $r_b$. (b) The path traced on the unit sphere $\mathbf{\hat{H}}'$ in the low-error regime. The upper hemisphere shows the slow initial spiral to the $x$-$y$ plane, followed by motion similar to Fig. 1e. (c) The correct binding configuration (right) next to the most commonly observed kinetic trap (left) for the case of a dimer with one (i) or two (ii) attracted monomers. (d) Growth of the number of dimers in the low-error regime with $\tau=300\tau_0$ and $q=12$ for initial dimer counts of 0 and 10. Each solid line is the average of 20 simulations of 1000 total dipoles. With more than 60.6 dimers, replication is the dominant channel for dimer production. Growth falls below exponential as the system exhausts free monomers.}  
	\label{fig:low_error}
\end{figure*}

The simplest method for suppressing $R_e$ is to lower monomer concentrations. However, this method is limited by monomer clusters in the bulk which lead to locally high monomer concentrations at the beginning of the mixing cycle. The stair-step structure seen in Fig. 2a shows bursts of dimers forming early in the mixing cycle due to these clusters. To achieve lower values of $R_e$ we propose a kinetic barrier between binding sites. The simplest barrier, a short range isotropic repulsion between colloids, is also among the most effective because the presence of a template negates the barrier. For monomers with no template, the repulsion will exponentially suppress binding events. 

Thus we find three general principles for high-quality replication: 1. The drive in the mixing phase should maximize $c$ -- dissolving all aggregates into well dispersed monomers and templates before a new replication phase starts. 2. The energy landscape should maximize $P_B(\infty)$. 3. Colloids should have an isotropic repulsion weaker than the template attraction but stronger than $k_BT$, with range greater than the binding length $r_b$ -- minimizing the random collision rate ($\propto R_e$) while only modestly reducing $R_r$. Although these conclusions are drawn from our investigation of self-replication in magnetic dimers, they extend to linear templates of any length and other forms of energy drive. 

We apply these principles to optimize the magnetic system. Isotropic repulsion naturally arises between wetted colloids and can be readily tuned through salt and pH, which makes this form of energy landscape manipulation experimentally attractive for any colloidal replication scheme. For spherical colloids larger than the screening length of the solution, electrostatic repulsion takes the form $q \exp{(-\kappa (r_{12}-\sigma))}$\cite{verwey}, where $q$ is an energy constant and $\kappa$ the inverse screening length. $q$ and $\kappa$ are chosen to satisfy the design principles just articulated. We set $\kappa$ to $0.2\sigma$ so that the kinetic barrier extends farther than $r_b$. $q$ is chosen in the range $k > q > k_BT$ to maximize $P_B$ by minimizing the two kinetic traps shown in Fig. 4c. Although it is not possible to simultaneously disfavor both traps relative to the binding configuration, a value of $q=0.6k$ achieves the optimal compromise (further discussed in Appendix D). We found that these potentials can be easily achieved for silicon-oxide coated magnetite particles in a neutral, low-salt solution as discussed in Appendix A. We also made system-specific alterations to the mixing field to mitigate the dimer explosion at the beginning of the mixing cycle. By initially spiraling slowly from the $z$ axis, the new form of $\mathbf{\hat{H}}'$ allows time for the monomer strings to dissolve before their binding sites are able to make contact. We additionally lengthened the mixing interval to improve $c$.

Figure 4d shows the long-term growth of dimers in this low-error regime. Fitting with Eq. (\ref{eq:Nd_full}) yields $X=0.0066\pm 0.001$. With such a low error rate the system is suitable for probing more delicate features of self-replication, such as environmental selection or structures more complex than dimers. 

\section{Conclusion}
We have shown that a system of self-replicable nanomaterials based on dipolar magnetic colloids produces high-quality autonomous replication. We considered autonomous self-replication as a stochastic process and quantified the effect of a finite probability that erroneous bonds form by introducing the parameter $X$. Exponential growth requires suppressing $X$ to values much less than 1. We identify three general design principles to minimize $X$ and apply them to create nearly pure exponential growth in a population of magnetic dimers.

There are many compelling avenues for future research with self-replicating magnetic colloids. Magnetic colloids have the unique advantage that they can be driven by magnetic fields in various environments, including living tissue. For this reason they are rapidly being adopted in a wide variety of medical applications\cite{kozissnik,duran}. Several current medical proposals may benefit from magnetic particles that can replicate patterns, such as tissue scaffolding\cite{tissue1,tissue2}. Achieving more complex self-replicating structures than the dimers discussed in here will require more sophisticated fabrication of magnetic colloids. Fortunately, recent experimental work with anisotropic magnetic particles suggests that more complex nonlinear clusters can be realized \cite{shifteddipole, janusmagnetic}. These advances in colloid synthesis, together with the rich and functional behavior of magnetic monomers in various types of magnetic fields, are expected to provide an avenue for self-replicating magnetic polymers and clusters. We believe further exploration of magnetic self-replication will be of both scientific interest and technological value.  

Acknowledgments. The authors thank Vladimir Kuzovkov, and Rebecca J. McMurray for helpful discussions. This work was supported by the Center for Bio-Inspired Energy Science (CBES), which is an Energy Frontier Research Center funded by the U.S. Department of Energy, Office of Science, Office of Basic Energy Sciences under Award Number DE-SC0000989. Numerical simulations were performed on TARDIS, which is a computer cluster financially supported by the Air Force Office of Scientific Research (AFOSR) under Award No. FA9550-10-1-0167.


\appendix

\section{Feasibility}
\subsection{Fabrication of particles}
One of the attractive features of the magnetic system in contrast to other theoretical proposals is its experimental accessibility. Magnetic colloids are routinely synthesized at diameters ranging from nanometers to microns. Fe$_3$O$_4$ (magnetite) is commonly used. At room temperature magnetite has monodomain sizes of 60 nm \cite{monodomain_size}. 

\begin{figure*}
\centering
	\includegraphics[width=2.0\columnwidth]{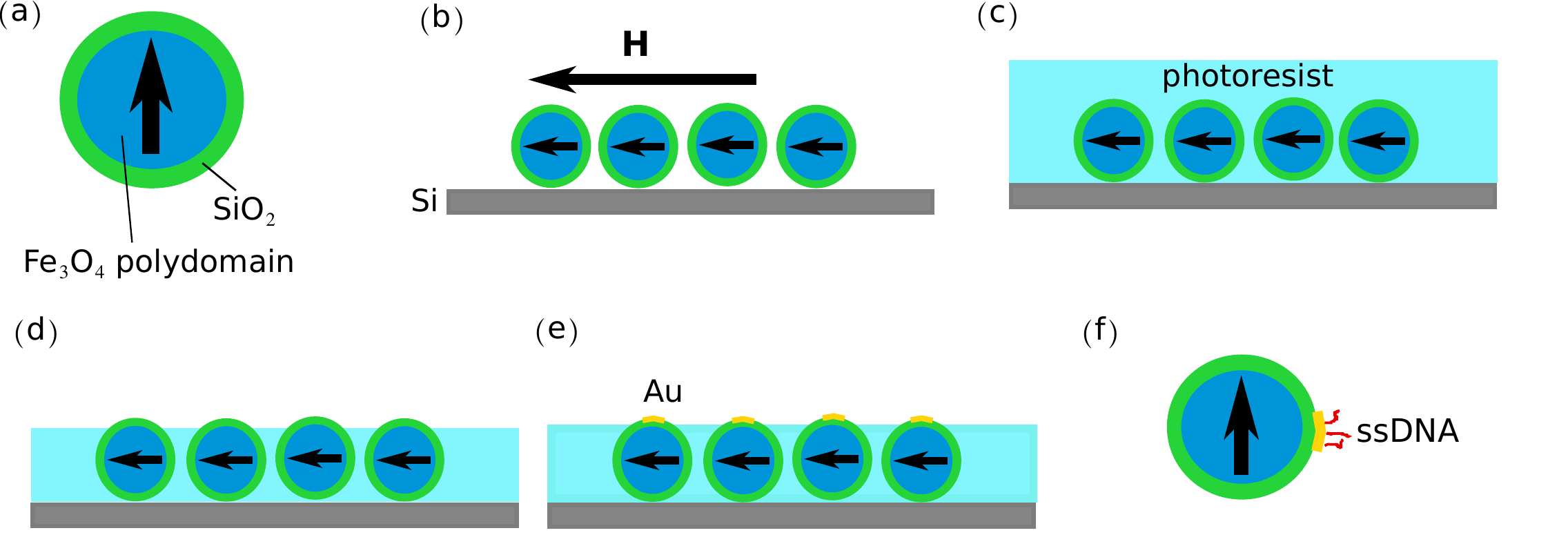}
	\caption{(Color online) A possible fabrication method, adapted from \cite{patch}. (a) Precursor colloids with a magnetite core coated by SiO$_2$ to a silicon surface. (b) The colloids are deposited on a silicon disc at low densities using a strong surfactant solution to prevent contact. Colloids will automatically assemble so that the surface presented to the interface is orthogonal to the dipole direction. An external field will assist this process and help prevent multi-layer aggregation.  (c) The disc is spin-coated with photoresist. (d) The photoresist is etched to slightly less than the colloid diameter. (e) Au is deposited on the exposed patches. (f) The photoresist is dissolved and colloids sonically removed and suspended in a solution with an appropriate surfactant. Colloids are finalized by self-complementary single-stranded DNA with an Au-linker at one terminus, which adsorbs to the gold patch.}
\end{figure*}

The only novel element for our particle system is the binding site perpendicular to the direction of the moment. We propose creating this site by assembly at an interface. Interactions between dipoles will automatically assemble them so their moments are in the plane of the interface. A large variety of methods have been used to create patchy particles at interfaces\cite{janus_overview}. Figure 6 shows one method for fabricating DNA patches using photoresist to protect most of the colloid surface\cite{patch}. This is applicable to colloid diameters between 100nm and 2$\mu$m. Our method differs from the original chiefly in the additional complication of long-range interactions. Maintaining a monolayer will require a low density of colloids and strong surfactant in the original solution. It may also require that an external field be applied while spin-casting the colloids onto the platter. If the resulting gold patches are too large for the replication scheme, they can be shrunk into compact crystals by heating to 700$^\circ$C and waiting for dewetting to occur\cite{shrinkpatch}. Since this is far above the Curie temperature for magnetite, the colloids would then be cooled in a strong external field to ensure their moments remained orthogonal to the patches.

Many alternative fabrication methods are possible. For example, McConnell and coworkers produced controllable patch sizes on silica particles as small as 103nm by embedding them in a polymer gel at an interface\cite{janus_gel}. In short, interested experimental groups have wide latitude in their choice of how to realize magnetic particles with lateral bonds.

\subsection{Achieving the quoted parameters}
Magnetite monodomains have magnetic moment per molecule of 4.1$\mu_B$, where $\mu_B$ is the Bohr magneton \cite{ferrites}. From this we find that monodomain particles of at least 9 nm will achieve $k=20k_BT$ at room temperature (25$^\circ$C), which is the value quoted for the low-error regime. Actual particles would need to be at least an order of magnitude larger, both for fabrication reasons and also to get permanent magnetic moments. Magnetic contact energies for larger particles can be reduced by thick nonmagnetic coatings or choosing a material with a lower remanence. Larger particles will also be multidomain and thus have reduced magnetization.

Achieving sufficiently low values of $\kappa$ requires low salt concentrations. For room-temperature water with a monovalent salt, an electrolyte concentration of $10^{-3}$ gives a screening length of 10 nm\cite{surfaces}, which would allow for a binding length of 5 nm while still satisfying $\kappa^{-1}=2r_b$. If the bond is realized with DNA, this would allow for strands of more than 15 base pairs. Still larger screening lengths can be achieved with less salt. 

Bousse and coworkers found the $\zeta$ potential for silicon oxide films in a pH 7 solution with $10^{-3}$ salt concentration to be roughly 60 mV\cite{bousse}. At this value, particles with a diameter of 100 nm will achieve $q \approx 200 k_BT$. This value is of the same order of magnitude as that found from direct experimental measurements of contact forces between colloids of different materials\cite{borkovec}. It is clear that achieving the requisite repulsion will not be difficult; if anything, it will be more challenging to have a sufficiently small potential. This can be done by decreasing pH and increasing salt, or using a coating with a lower $\zeta$ value in the desired solution.

In sum: the parameters used in this work can be readily realized for particles consisting of a magnetite core and silicon coating in a neutral solution with low salt. In fact, it may be preferable to work with larger values of $k$ and $q$ and smaller values of $r_b$ relative to $\sigma$. In simulations we are limited by computing time that scales with the square of potential strength, but we expect stronger potentials to produce clearer replication data. 


\section{Methods}

\subsection{Kinetic Monte Carlo with rotations}
Simulations in this paper use the kinetic Monte Carlo (KMC) method developed by Jha et al \cite{jha}. This approach is well-suited to modeling non-equilibrium mechanics for long time scales, especially if the system, like ours, does not form large, compact clusters. A more detailed derivation of the method with rotations and dimers has already been presented \cite{us}. In brief, the method consists of attempting to move a particle a distance $a$ in a random direction, calculating the change in energy the move would cost, and accepting or rejecting the move with probability
\begin{align}
	P=\frac{1}{1+e^{\Delta U}}
\end{align}
where $\Delta U$ is the change in potential the move would cause in units of $k_BT$. Once each particle has attempted its move the KMC sweep is complete. For spherical monomers we find the correct ratios of translational to rotational diffusion using the Einstein relation and the well-known drag coefficients for spheres:
\begin{align}
	\frac{\langle r^2 \rangle}{\langle \theta^2 \rangle}=\frac{6Dt}{6D_rt}=\frac{\mu}{\mu_r}=\frac{\pi \eta \sigma^3}{3 \pi \eta \sigma}=\frac{\sigma^2}{3}
\end{align}
where $\eta$ is the viscosity of fluid, $\mu$ the mobility of the spheres, and $\sigma$ the diameter of the monomer. We choose the rotational step size to be
\begin{align}
	\phi = \frac{a}{\sigma}\\\nonumber
\end{align}
so to preserve the correct diffusion ratios the probability of rotating the monomer must be three times the probability of translating it.

We can find appropriate step sizes for the dimers by comparing their diffusion rates to the rates for spherical particles. Let $a_1$ be the translational step size for dimers along their long axes, and $a_2,\; \phi_2$ the translational and rotational step sizes in the perpendicular directions. We then write
\begin{equation} 
	\begin{array}{rcl}
		a_1^2 & = & a^2 \frac{2D_1}{6D}\\[.2cm]
		a_2^2 & = & a^2 \frac{2(d-1)D_2}{2dD}\\[.2cm]
		\phi_2^2 & = & \phi^2 \frac{2(d-1)D_{r2}}{2dD_r}
	\end{array}
\end{equation}
The ratios of the drag coefficients of a dimer to those for individual spherical particles are known exactly\cite{torre}. From these we determine $a_1 = 0.51a$, $a_2 = 0.68a$, and $\phi_2=0.42 \phi$. We have not calculated the step size for rotation about the long axis because the dimer members rotate independently. Whenever we would attempt to rotate the dimer in this direction, we will instead attempt rotating the two particles each by an angle $\pm \sfrac{\phi}{\sqrt{3}}$, where the factor $\sfrac{1}{\sqrt{3}}$ arises because the rotations are confined to one dimension.

In sum, during a sweep we try translating each monomer by $a$ 12.5\% of the time and rotating by $\phi$ 37.5\% of the time in a random direction or about a random axis, respectively. For dimers we will attempt the two translational moves  12.5\% of the time (for each one) and the two rotational moves with probability 37.5\% of the time (for each one). Because eight sweeps are required to diffuse properly in each degree of freedom, the time scale for each sweep is $\Delta t = \frac{a^2}{96D}$. Let $\tau_0$ be the time it takes a particle to diffuse a distance equal to its diameter, $\sfrac{\sigma^2}{6D}$, so that $\tau_0 = \pi \eta_s \sigma^3/2k_BT$, where $\eta_s$ is the solvent viscosity. Then we find
\begin{align}
	\frac{\Delta t}{\tau_0}= \frac{1}{16} \left( \frac{a}{\sigma} \right) ^2
\end{align}

\subsection{Definitions of constants}
There are five relevant potentials to simulate. The first is the dipole-dipole potential that controls magnetic interactions,
\begin{equation}
	U_\mathrm{dipole}=
	\begin{cases}
	k \left( \frac{\sigma}{r} \right)^3 \left ( \hat{\mathbf{\mu}}_1 \cdot \hat{\mathbf{\mu}}_2 - 3 ( \hat{\mathbf{\mu}}_1 \cdot \hat{\mathbf{r}} )( \hat{\mathbf{\mu}}_2 \cdot \hat{\mathbf{r}} ) \right )
	& r < 6 \sigma\\
	0
	& r \ge 6 \sigma\\
	\end{cases}
	\label{eq:dipolepotential}
\end{equation}
where $\mathbf{\mu}_i$ is the direction of the ith dipole. The energy scale $k$ is given in SI units by
\begin{equation}
	k=\frac{\mu_0 \mu^2}{4 \pi \sigma^3}
	\label{eq:k}
\end{equation}
with $\sigma$ the colloid diameter. Note that since $\mu$ scales with colloid volume, so does $k$.

The second potential is the external magnetic field:
\begin{equation}
	U_{\mathrm{external}} = H_\mathrm{ext} \hat{\mathbf{H}}_\mathrm{ext}\cdot \hat{\mathbf{\mu}}
	\label{eq:external}
\end{equation}
If the external field is $\mathbf{B}$, then $H_\mathrm{ext} = \mu B$ in SI units. It takes on the values $H$ and $H'$ over the course of the replication cycle. Note that $H_\mathrm{ext}$ has dimensions of energy, not magnetic induction.

The third potential is a hard-sphere interaction. In kinetic Monte Carlo simulations hard-sphere potentials are implemented as a rule preventing two particles from approaching closer than $\sigma$. 

The fourth potential is the interaction between binding sites. For the simple magnetic system, this is implemented as an infinite well with radius $2r_b$ and a minimum at $r_b$. Once two binding sites are close enough, they bind permanently with perfect rigidity at a separation of $r_b$. The two colloids members of the dimer are free to twist around the axis of this bond. The dimer itself moves as a unit in simulation; no actual potential is attributed to the bond.

The final potential comes from electrostatic repulsion between wetted colloids. This potential has the form\cite{verwey} 
\begin{equation}
	U_q = q e^{-\kappa(r-\sigma)}
	\label{eq:Uq}
\end{equation}
Here,  $q = \pi \epsilon \sigma \zeta^2$\cite{goodwin}, where $\epsilon$ is the permittivity of the solution and $\zeta$ is a characteristic electrical potential that depends strongly on pH. $\kappa$ scales with the root of the ion concentration.

In all cases we chose energy units such that the thermal energy $k_BT$ is one.

\subsection{Parameters}
For the simple ($q=0$) dipole system, two kinds of simulation $S$ and $S'$ were used to improve statistical precision. In $S$ whenever two binding sites are close enough they form a permanent, perfectly rigid dimer as described above. These simulations have 2500 colloids of which 20 are initially bound in dimers. In $S'$ no dimer was formed when binding sites are within $2 r_b$. Instead, the binding sites are deactivated. However we still ``ticked" the dimer counter for the system. These simulations began with 2490 free colloids and no dimers. Thus $S'$ contained no templates and showed only dimers formed erroneously. Comparing the two systems allowed greater precision in calculating the replication rate $R_r$ and the error rate $R_e$. 

The plots in Fig. 3 are based on 18 different values of $\tau$. For each value of $\tau$, 50 simulations each of $S$ and $S'$ were run. In each simulation the step size was 0.1$\sigma$. This is a "proper" step size  (defined in ref. \cite{rui}) except for particles in the long ferromagnetic strings, whose central members were at the bottom of a well with depth lower than $80k_BT$. However, since these strings were repeatedly dissolved on timescales much shorter than their diffusion, we were willing to accept their artificially slow motion in exchange for simulating longer time scales.

The other parameters for the simple dipole system are $k=20 k_BT$; $r_b = 0.05\sigma$; $\phi=0.00226$; $H=5 k_BT$; $H'=1000 k_BT$; and $q=0$, corresponding to negligible electrostatic interactions. $\mathbf{H}$ always points along the $z$ axis. For the direction of the mixing field $\mathbf{H'}$, we need a continuous function that samples all points with close to uniform distribution. We chose a vector whose (unnormalized) components are defined as follows:
\begin{align}
	F_x(t')&=\cos{\omega\frac{t;}{t_s}}\notag\\
	F_y(t')&=\sin{\omega\frac{t;}{t_s}}\notag\\
	F_z(t')&=
	\begin{cases}
		1-\sin{\pi\frac{t'}{t_s}} & 0 \leq t' \leq 0.5\tau_s\\
		-1+\sin{\pi\frac{t'}{t_s}} & 0.5 \tau_s \leq t' \leq \tau_s\\
		-1-\sin{\pi \frac{t'}{t_s}} & \tau_s \leq t' \leq 1.5\tau_s\\
		1+\sin{\pi\frac{t'}{t_s}} & 1.5\tau_s \leq t' \leq  2\tau_s\\
	\end{cases}
	\label{eq:F}
\end{align}
where $t'$ is the time measured from the start of the mixing period. The components of $\mathbf{M}'$ are
\begin{align}
	H'_x & = H'F_x(1-F_z^2)^{\sfrac{1}{2}}\notag\\
	H'_y & = H'F_y(1-F_z^2)^{\sfrac{1}{2}}\notag\\
	H'_z & = H'F_z
	\label{eq:Mp}
\end{align}

This function oversamples points near the $z=0$ plane. Since function changes direction fastest at $z=0$, the uneven distribution improves cluster dissolution. We imposed the mixing field for $12 \tau_0$ each cycle with $t_s=0.375\pi$ and $\omega=2.1$.

To find a regime with low errors, we set $q=12$, $\phi=0.001$, $\kappa=5\sigma^{-1}$, $H=4$, and $H'=200$.

We left the other parameters unchanged. However, we altered the mixing field direction so that
\begin{align}
	F_{z2}(t')&= 
	\begin{cases}
		\cos\left(\frac{0.5\pi}{a}\frac{t'}{\tau_s} \right) & t' \le a \tau_s\\
		F_z \left( \frac{t'}{t_s} \right)\ & t' > a \tau_s\\
	\end{cases}
	\label{eq:F2}
\end{align} 
with values $\omega=2\sqrt{2}$, $t_s=2\tau_0$, and $a=20.25$. The first part of the function produces a very gradual motion towards the x-y plane early in the mixing cycle. This prevents an initial burst of erroneous dimers caused by rotating the colloids in long strings too quickly. The reduced speed of the dissolving field helps prevent errors later in the cycle, at the cost of less efficient dissolution. Overall we lengthened the mixing time to $t_m=49\tau_0$.

Figure 2a is based on data from 10 simulations that begin with 500 monomers randomly distributed and no dimers. All other parameters are as in the simple dipole system ($k=20 k_BT$; $r_b = 0.05\sigma$; $d=0.00226$; $H=5 k_BT$; $H'=1000 k_BT$; $q=0$; $\hat{\mathbf{H}}'$ given by Eq. (\ref{eq:Mp})). The system is allowed to evolve in a KMC simulation for $\tau=100\tau_0$, then mixed for $t_m=12\tau_0$ and allowed to evolve for a second interval $\tau=100\tau_0$. Colloids are counted as ``free'' if their centers are more than $1.5\sigma$ apart. The number of free colloids in the first interval forms the dashed blue line of Fig. 2a (random initial configuration), while the second interval forms the solid green line (after mixing field).

Figure 2b is drawn from 10000 simulations of a single dimer with two adsorbed monomers. The dimer and monomers begin oriented in the binding configuration as in Fig. 1d(ii). The two monomer positions are on average centered directly above the two dimer positions, but randomly varied by $0.5r_b$ in all directions at the beginning of each simulation. The monomer dipole moments begin oriented strictly parallel to the corresponding moments of each member of the dimer and anti-parallel to each other. Aside from this constraint, the binding sites are oriented randomly. All parameters are as quoted in the preceding paragraph. The system evolved in a KMC simulation until either the two monomers bind or $100\tau_0$ has passed. Each binding event constitutes a data point. The solid green line in Fig. 2b shows the cumulative sum of bindings over time normalized by 10000. The exponential curve shown by the blue dotted line is not determined by any detailed analysis. It is included only to illustrate that $P_b$ falls progressively  short of an exponential distribution, with the implications which are discussed in the main text.

\newpage
\section{Data analysis}
\subsection{Models for $N_d(t)$}
 As given in the main text, the exact solution for dimer expectation value is
\begin{equation}
	 N_d(t) =\frac{N}{4(1-X)}\tanh \left ( \frac{R_r}{2}t+b \right) +\frac{N(1-2X)}{4(1-X)}
	 \label{eq:Nd1_full}
\end{equation}
where $X$ is the ratio $\sfrac{R_e}{R_r}$, and $b$ is chosen to match the initial number of dimers, $N_d(0)$. In the limit $N \gg N_d$ this simplifies to
\begin{equation}
	 N_d(t) = \left ( N_d(0)+\frac{N}{2} X \right ) e^{R_rt}-\frac{N}{2} X
	\label{eq:Nd1}
\end{equation}
If, as in $S'$, the dimers do not act as templates or otherwise affect dimer formation, then $R_r=0$ and the solution to Eq. (1) is
\begin{equation}
	 N_d(t) = \frac{N}{2} \left( 1-e^{-R_et} \right)
	\label{eq:Nd1'}
\end{equation}

\subsection{Statistical methods used}
In each simulation run we count ``hits", where each hit is a dimer forming. A logical form for the probability is the Poisson distribution. Since we begin each system with 10 dimers, the distribution is shifted by 10:
\begin{align}
	P(N_d, t)=\frac{(\mu(t)-10)^{N_d-10}}{(N_d-10)!}e^{\mu(t)-10}
	\label{eq:Poisson}
\end{align}
where $\mu$ is the mean value calculated from one of Eqs. (\ref{eq:Nd1_full}), (\ref{eq:Nd1}), or (\ref{eq:Nd1'}). For a given set of parameters, the total probability of observing a particular set of runs is
\begin{align}
	P_O=\prod_{S\;\mathrm{runs}}\prod_t 						P(N_d(\mathrm{run},t),t)
	\label{eq:P_T}
\end{align}
We define $M$ as $-\ln{P_O}$, or
\begin{align}
	M=-\sum_{\mathrm{runs}}\sum_t& (N_d(\mathrm{run},t)-10)\ln{\mu(t)-10}\notag\\
	&-\ln{(N_d(\mathrm{run},t)-10)!}+\mu(t)-10
	\label{eq:M}
\end{align}
To maximize $P_O$ is equivalent to minimizing $M$. To do this we use a fixed-step grid search in parameter space. Step sizes in the parameters $x^i$ are chosen to be less than $(\partial_{x^i}^2 M)^{-\sfrac{1}{2}}$, i.e. fine enough that numerical errors are much less than the minimum standard errors for the parameters.

In order to develop some distribution of values for the parameters, we use a ``run-replacement" bootstrapping method. We treat the set of runs of each simulation type as a parent distribution. If there are $n$ unique runs in the distribution, we select $n$ runs from it. The same run may be selected more than once. We then find the best-fit parameters for the selection by grid search. By repeating this process 200 times we develop a distribution for each parameter. Standard errors are estimated from the standard deviations of the resulting distributions.  

\subsection{The simple dipole system ($q=0$)}
For the simple ($q=0$) dipole system, two kinds of simulation $S$ and $S'$ were used to improve statistical precision. In $S$ whenever two binding sites are close enough they form a permanent, perfectly rigid dimer as described above. These simulations have 2500 colloids of which 20 are initially bound in dimers. In $S'$ no dimer was formed when binding sites are within $2 r_b$. Instead, the binding sites are deactivated. However we still ``ticked" the dimer counter for the system. These simulations began with 2490 free colloids and no dimers. Thus $S'$ contained no templates and showed only dimers formed erroneously. Comparing the two systems allowed greater precision in calculating the replication rate $R_r$ and the error rate $R_e$. 

For each simulation run, we averaged the number of dimers over the course of each complete replication cycle. Thus, each run has a number of data points equal to the number of cycles completed during the run. For runs of the $S$ (normal) type, which begin with 10 dimers, we use a model based on Eq. (\ref{eq:Nd1}):
\begin{align}
	\mu(t)=\left(10+C+\frac{N}{2}X\right)e^{R_rt}-\frac{N}{2}X
	\label{eq:S}
\end{align}
where $N$ is the total number of dipoles in the system and the constant $C$ allows for some difference between the replication rate in the first cycle (which is computer randomized) and subsequent cycles (which are randomized by the drive). For $S'$, we use a model based on Eq. (\ref{eq:Nd1'}):
\begin{align}
	\mu'(t)=\frac{N}{2}\left(1-e^{-R_et}(1-\frac{2C'}{N})\right)
	\label{eq:S'}
\end{align}
where once again $C'$ allows for a different rate on the first cycle. The value of $M$ depends on the four parameters $R_r$, $R_e$, $C$ and $C'$. 

An important question is to what order the curves in Figure 2 are exponential. We confirm growth to cubic order using a different kind of fitting than that used above. The expansion of (\ref{eq:Nd1}) is
\begin{align}
	N_d(0)+\left( N_d(0)+\frac{NR_e}{2R_r} \right)\left(R_rt+\frac{R_r^2t^2}{2}+\frac{R_r^3t^3}{6}+...\right)
	\label{eq:Ndexpand}
\end{align}
The two independent parameters $R_e$ and $R_r$ make the linear and quadratic moments of this expansion independent, and therefore to fit the exponent up to the quadratic term we use the model $y_2(t)=at^2+bt+c$. The cubic term, however, is not:
\begin{align}
	y_3(t)=c+bt+at^2+\frac{2a^2}{3b}t^3
	\label{eq:y3}
\end{align}
We compare the coefficient of determination for the quadratic and cubic fits to runs in $S$ and find that the ternary term dramatically improves the quality of all fits, from $R^2 \in [0.84,\,0.91]$ to $R^2 \in [0.97,\,0.99]$. Higher order terms cannot significantly improve fitting quality, so we cannot confirm exponential growth beyond the third term. 

\subsection{Low-error regime ($q \neq 0$)}
The low-error regime provides a much clearer signal with a smaller sample. We used only 20 runs of 1000 particles to build a distribution for each case $N_d(0)=0$ and $N_d(0)=10$. The smaller simulation sizes make exhaustion effects significant ($N_d > 0.1 N$). Consequently, we use the exact solution Eq. (\ref{eq:Nd1_full}) as the mean in Eq. (\ref{eq:M}). In this case, no constant offset was necessary to allow for different replication rates in the first cycle, as the dimer formation rate per cycle was too small for this effect to be significant.

\subsection{Fitting the replication rate}
To find $R_r(\tau)$ we minimized the reduced chi-squared measure
\begin{equation}
\chi^2 = \frac{1}{n-2}\sum_{i=1}^n \left( \frac{(r(\rho, \xi, \tau_i) - R_{r,\,i}}{s_{i}} \right)^2
	\label{eq:chisquared}
\end{equation}
where $r(\rho, \xi, \tau)$ is the model for $R_r(\tau)$ presented in Section III with the distributions for $P_B$ and $\eta_f$ found in Figure 3, and $R_r$ and $s$ are the observed mean and deviation for a given value of $\tau$, as described above. We find the reduced chi-squared value between 0.86 and 0.9 for a wide parameter space so long as $0.04  < \eta_0 \rho \tau_0 < 0.06$ and $\xi = 0.0024 (\eta_0 \rho \tau_0) ^{-2}$. This indicates both that there is only one effective degree of freedom in the fitting region and that bootstrapping overestimates the errors for replication rates.



\newpage
\section{Kinetic trapping}
\begin{figure}[h]
\centering
	\includegraphics[width=.95\columnwidth]{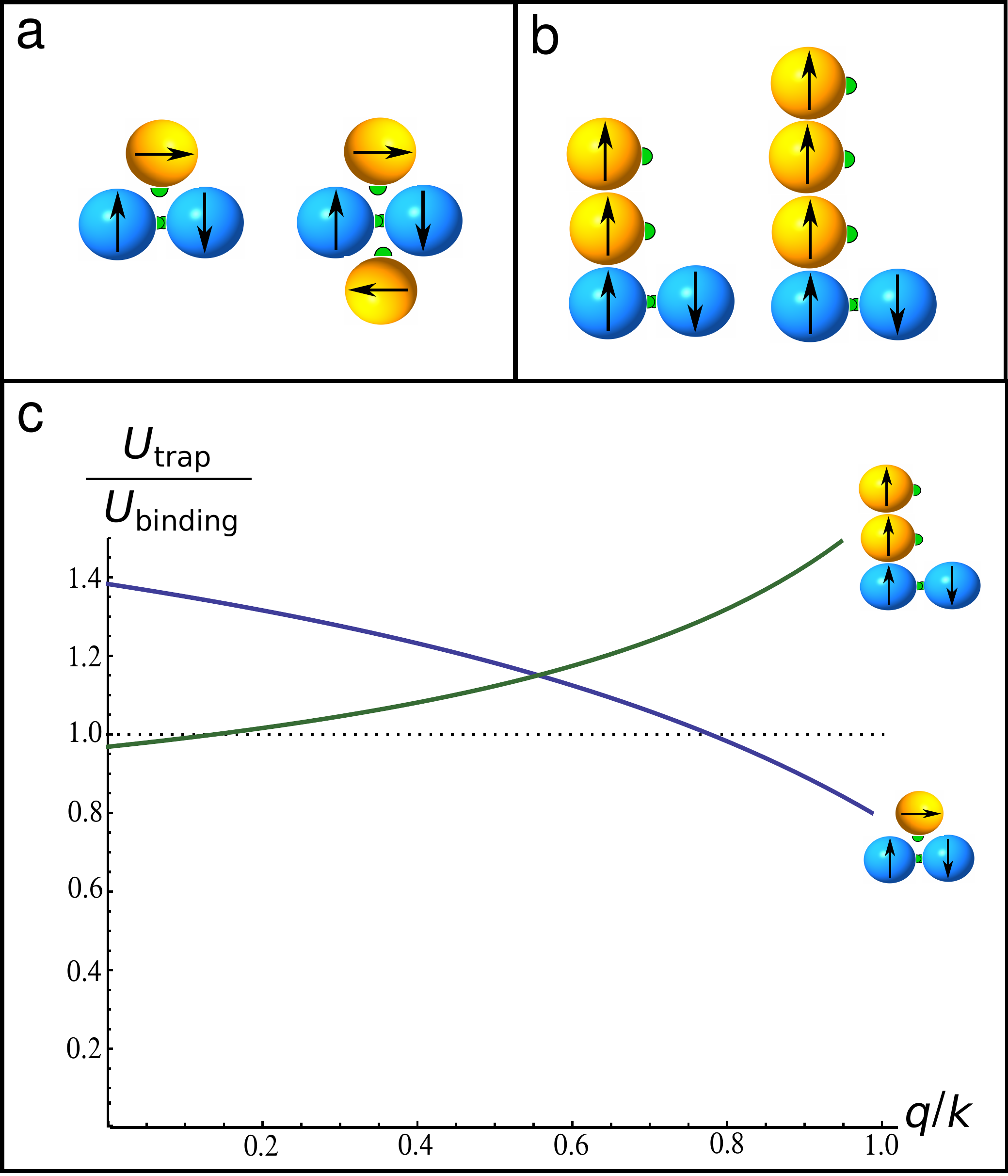}
	\caption{(Color online) Kinetic trapping. (a) The triangle trap family, shown with one and two particles. (b) The line trap family, shown with two or three particles. (c) Trapping energies normalized by the binding energies of the corresponding configuration. A value greater than one indicates that the trap is energetically favorable. In this plot, $H=0.25k$}
\end{figure}
Unlike a molecular or patchy colloidal replicator, magnetic dipole potentials are smooth and long-range. This has the advantage of allowing objects to replicate efficiently at very low densities, but the disadvantage of creating minima in configuration space which will prevent replication. The barriers between local minima are not too strong for the parameters used in this paper, and we frequently observe transitions among them. However, if one of the non-binding minima is significantly lower in energy than the correct configuration it can greatly impede replication. Here we briefly discuss a parameter choice that mitigates this problem.

We classify the undesirable minima into two families as shown in Fig. 6. The first is the triangular trap (Fig. 6a). Here a monomer adsorbed on a dimer does not remain one member or the other of the dimer but sits between the two. This configuration blocks other monomers from adsorbing on that side of the dimer. Because it is more compact than the correct configuration, this configuration is disfavored by increasing $q$. For a single monomer, the energy of the triangle trap (with interactions between dimer members excluded) is $U_\mathrm{triangle}=-2.6k-H+2q$, while the correct configuration has energy $-1.8k-H+q$. The second undesirable configuration is the line trap, in which the second monomer on a dimer adsorbs onto the first (Fig. 6b). This trap has energy $U_\mathrm{line} = -4k -2H +2q$, while the correct (square) configuration with two monomers has energy $-4.6k +3q$.

Fig. 6c plots the energies of the two traps normalized by the energy of the corresponding correct configuration. With the values of $H$ used previously it is impossible to simultaneously disfavor both traps. The best compromise can be found at $q=0.6k$, at which the energies of the two traps are both slightly larger than the desired configuration energies.


\begin{thebibliography}{99}

\bibitem{1986paper} G. von Kiedrowski, Angew. Chem. Int. Ed. \textbf{25,} 932 (1986).

\bibitem{joyce} N. Paul and G. F. Joyce, Proc. Natl. Acad. Sci. U.S.A. \textbf{99,} 12733 (2002).

\bibitem{rnaenzyme} T. A. Lincoln and G. F. Joyce, Science \textbf{323,} 1229 (2009).

\bibitem{jacs2002} R. Issac and J. Chmielewski, J. Am. Chem. Soc. \textbf{124,} 6808 (2002).

\bibitem{rebek} T. Tjivikua, P. Ballester, and J. Rebek Jr., J. Am. Chem. Soc. \textbf{112,} 1249 (1990).

\bibitem{vidonne} A. Vidonne and D. Philp, Eur. J. Org. Chem. \textbf{5,} 593 (2009).

\bibitem{chaikinsm} M. E. Leunissen, R. Dreyfus, R. Sha, T. Wang, N. C. Seeman, D. J. Pine, and P. M. Chaikin, Soft Matter \textbf{5,} 2422 (2009).

\bibitem{chaikin1} T. Wang, R. Sha, R. Dreyfus, M. E. Leunissen, C. Maass, D. J. Pine, P. M. Chaikin, and N. C. Seeman, Nature \textbf{478,} 225 (2011).

\bibitem{us} R. Zhang, J. M. Dempster, and M. Olvera de la Cruz, Soft Matter \textbf{10,} 1315 (2014).

\bibitem{brenner} Z. Zeravcic and M. Brenner, Proc. Natl. Acad. Sci. U.S.A. \textbf{111,} 1748 (2014).

\bibitem{angew} R. Zhang, D. A. Walker, B. A. Grzybowski, and M. Olvera de la Cruz, Angew. Chem. Int. Ed. \textbf{53,} 173 (2014).

\bibitem{otto} J. M. A. Carnall, C. A. Waudby, A. M. Belenguer, M. C. A. Stuart, J. J.-P. Peyralans, and S. Otto, Science \textbf{327,} 1502 (2010).

\bibitem{DNAjanus} H. Xing, Z. Wang, Z. Xu, N. Y. Wong, Y. Xiang, G. L. Liu, and Y. Lu, ACS Nano \textbf{6,} 802 (2012).

\bibitem{janus_overview} A. Walther and A. H. E. M\"{u}ller, Chem. Rev. \textbf{113,} 5194 (2013).

\bibitem{trusov} L.A. Trusov, A. V. Vasiliev, M. R. Lukatskaya, D. D. Zaytsev, M. Jansen, and P. Kazin, Chem. Commun. \textbf{50,} 14581 (2014).

\bibitem{halsey} J. E. Martin, J. Odinek, T. C. Halsey, and R. Kamien, Phys. Rev. E \textbf{57,} 756 (1998).

\bibitem{jha} P. K. Jha, V. Kuzovkov, and M. Olvera de la Cruz, ACS Macro Lett. \textbf{1,} 1393 (2012).

\bibitem{sm} See Supplemental Material at [URL will be inserted by publisher] for two simulation movies and their descriptions. 

\bibitem{verwey} E. J. W. Verwey and J. Th. G. Overbeek, \textit{Theory of the Stability of Lyophobic Colloids} (Elsevier, New York, 1948).

\bibitem{kozissnik} B. Kozissnik and J. Dobson, MRS Bull. \textbf{38,} 927 (2013).

\bibitem{duran} J. D. G. Dur\' an, J. L. Arias, V. Gallardo, and A. V. Delgado, J. Pharm. Sci. \textbf{97,} 2948 (2007).

\bibitem{tissue1} A. Ito and M. Kamihira, Prog. Mol. Biol. Trans. Sci \textbf{104,} 355 (2011).

\bibitem{tissue2} K. Buyukhatipoglu, R. Chang, W. Sun, and A. M. Clyne, Tissue Eng. C: Methods \textbf{16,} 631 (2010).

\bibitem{shifteddipole} S. Sacanna, L. Rossi, and D. J. Pine, J. Am. Chem. Soc. \textbf{134,} 6112 (2012).

\bibitem{janusmagnetic} J. Yan, M. Bloom, S. C. Bae, E. Luijten, and S. Granick, Nature \textbf{491,} 578 (2012).  

\bibitem{monodomain_size} A. Aharoni and J. P. Jakubovocs, IEEE Trans. Mag. \textbf{24,} 1892 (1988).

\bibitem{patch}Z. Bao, L Chen, M Weldon, E Chandross, O. Cherniavskaya, Y. Dai, and J. B.-H. Tok, Chem. Mater. \textbf{14,} 24 (2002).

\bibitem{shrinkpatch} Y. Lu, H. Xiong, X. Jiang, and Y. Xia, J. Am. Chem. Soc. \textbf{125,} 12724 (2003).

\bibitem{janus_gel} M. D. McConnel, M. J. Kraeutler, S. Yang, and R. J. Composto, Nano Lett. \textbf{10,} 603 (2010).

\bibitem{ferrites}B. Viswanathan, in \textit{Ferrite Materials: Science and Technology}, edited by B. Viswanathan and V. R. K. Murthy (Narosa, New Delhi, 1990).

\bibitem{surfaces}  D. Myers, \textit{Surfaces, Interfaces, and Colloids: Principles and Applications}(Wiley-VCH, New York, 1999).

\bibitem{bousse} L. Bousse, S. Mostarshed, B. v. d. Shoot, N. F. de Rooij, P. Gimmel, and W. G\"{o}pel, J. Colloid Interface Sci. \textbf{147,} 22 (1991). 

\bibitem{borkovec}F. J. Montes Ruiz-Cabello, G. Trefalt, P. Maroni, and M. Borkovec, Phys. Rev. E \textbf{90,} 012301 (2014).

\bibitem{torre}B. Carrasco and J. Garc\'{i}a de la Torre,   Biophysical Journal \textbf{75,} 3044-3057 (1999).

\bibitem{goodwin} J. Goodwin, \textit{Colloids and Interfaces with Surfactants and Polymers} (Wiley, London, 2009).

\bibitem{rui} R. Zhang, P. K. Jha, and M. Olvera de la Cruz, Soft Matter \textbf{9,} 5042-5051 (2013).

\end{thebibliography}
\end{document}